\documentclass[letterpaper, 10 pt, conference]{ieeeconf}
\usepackage{geometry}

\IEEEoverridecommandlockouts 
\usepackage{graphicx}
\usepackage{color}
\usepackage{subcaption}
\usepackage{amsmath}
\usepackage{amssymb} 
\usepackage{tabularx} 
\usepackage{flushend}
\usepackage{geometry} 
\usepackage{bm} 

\title{\LARGE \bf
Trajectory Tracking Control of a Flexible Spine Robot, \\With and Without a Reference Input
}

\author{Andrew P. Sabelhaus*$^{1\dagger}$, Shirley Huajing Zhao$^{1}$, Mallory C. Daly$^{1\dagger}$, Ellande Tang$^{1}$,  Edward Zhu$^{1}$,\\ Abishek K. Akella$^{2}$, Zeerek A. Ahmad$^{3}$, Vytas SunSpiral$^{4\dagger}$, Alice M. Agogino$^{1}$
\thanks{\it * corresponding author.}
\thanks{$\dagger$Authors with the NASA Ames Intelligent Robotics Group and the Dynamic Tensegrity Robotics Lab, Moffett Field CA 94035, USA}%
\thanks{$^{1}$Authors with
        the Department of Mechanical Engineering, 
        University of California Berkeley, USA
        {\tt\small apsabelhaus@berkeley.edu}}%
\thanks{$^{2}$A.K. Akella is with Levant Power Corp., 475 Wildwood Ave, Woburn MA 01801, USA
        {\tt\small akakella@berkeley.edu}}%
\thanks{$^{3}$Z.A. Ahmad is with Velo3D Inc.,
        1001 Belford Dr., San Jose, CA 95132, USA
        {\tt\small zeerekahmad@gmail.com}}%
\thanks{$^{4}$V. SunSpiral is with SGT Inc., Greenbelt, MD 20770, USA, working at
        NASA Ames Research Center, Moffett Field CA 94035, USA
        {\tt\small vytas.sunspiral@nasa.gov}}%
}

\begin{document}

\maketitle
\thispagestyle{empty}
\pagestyle{empty}

\begin{abstract}

The Underactuated Lightweight Tensegrity Robotic Assistive Spine (ULTRA Spine) project is an ongoing effort to develop a flexible, actuated backbone for quadruped robots.
In this work, model-predictive control is used to track a trajectory in the robot's state space, in simulation.
The state trajectory used here corresponds to a bending motion of the spine, with translations and rotations of the moving vertebrae.
Two different controllers are presented in this work: one that does not use a reference input but includes smoothing constrants, and a second one that uses a reference input without smoothing.
For the smoothing controller, without reference inputs, the error converges to zero, while the simpler-to-tune controller with an input reference shows small errors but not complete convergence.
It is expected that this controller will converge as it is improved further.




\end{abstract}

\section{INTRODUCTION}

The Underactuated Lightweight Tensegrity Robotic Assistive Spine (ULTRA Spine) is an ongoing project to develop a flexible, actuated spine for quadruped robots.
This involves creating a control system for the spine's cables, so that it can perform the necessary bending motions.
The spine is made out of a tensegrity ("tensile-integrity") system, where cables in tension hold the spine's vertebrae apart, and where the lengths of these cables are adjusted as inputs.

\begin{figure}
	\centering
	\includegraphics[width=0.9\columnwidth]{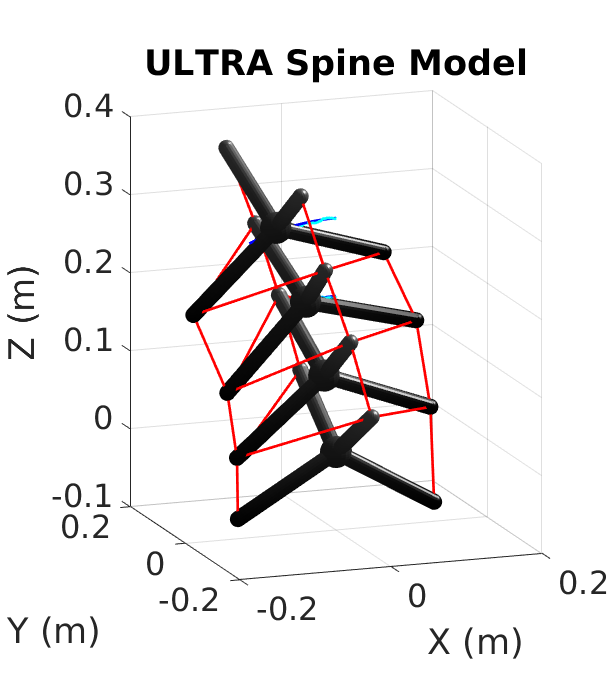}
	\caption{Trajectory-tracking control for the flexible backbone robot (ULTRA Spine), mid-simulation, for a uniaxial bending trajectory, using the method presented in \cite{Sabelhaus2017a}. The rigid bodies (vertebrae) of the spine are in gray, cables in red, and the target trajectory for the top spine vertebra is in blue. This work uses a point-mass dynamics model, so the rigid vertebra bodies are for visualization only.}
	\label{fig:ultra_spine_mid-bend_mpc}
\end{figure}

This work considers two similar models of the ULTRA Spine, and applies trajectory tracking controllers to each of them (Fig. \ref{fig:ultra_spine_mid-bend_mpc}.)
The first controller is presented in more depth in recent work by the authors \cite{Sabelhaus2017a}, while the second controller with input constraints is ongoing work and is presented here for the first time.
That paper includes more background on the motivations and choices made for this controller.

\section{CONTROLLER FORMULATIONS}\label{sec:controller}

Two different a model-predictive controllers were used to track a reference trajectory $\xi^{ref}$, which corresponded to counterclockwise bending of the spine.

This work uses a Model-Predictive Control (MPC) law for multiple reasons.
First, there are inherent constraints on the dynamics of this system: the rest lengths of the springs cannot be negative (the springs can only compress down to a certain length), and the vertebrae of the system should not contact each other.

In addition to these two constraints, there is an additional requirement (not considered here) on the cable tensions: the cables cannot ''push'', e.g. tensions must be non-negative; however, this is now included in the dynamics derivation so is not considered as a constraint. 

However, most importantly, constraints were used here to improve the quality of the linearization.
Prior controllers became unstable due to the poor linearizations and the rapid movements that were created.
By constraining and penalizing the amount of movement of the spine, the controllers create feasible movements.

\subsection{Model-Predictive Control Overview}

At each timestep $t$, model-predictive control solves a constrained finite-time optimal control problem (CFTOC), generating the sequence of control inputs $U_{t\rightarrow t+N|t} = \{u_{t|t}, ..., u_{t+N|t}\}$, with a window of $N$.
The notation $t+k|t$ represents a value at the timestep $t+k$, as given or predicted at timestep $t$ (\cite{Borrelli2003}, Ch. 4.)
Then, the first input $u_{t|t}$ is applied to the system, and the simulation advances to timestep $t+1$, and the problem repeats.
Note that no terminal costs or constraints are included here, and thus stability can only be experimentally concluded, not proven.

The two different CFTOCs for the controllers are listed below.
The formulation of this optimization problem is the only difference between the controllers, alongside the fact that they were applied to slightly different models of the ULTRA Spine.

\subsection{Controller with smoothing, without a reference input}

The first controller used a four-vertebra, three-dimensional model of the ULTRA Spine, and solved the optimization problem below for each step of MPC.
Note that for this controller, no corresponding input to the state trajectory was available, so smoothing constraints had to be applied.
These are explained more in-depth in \cite{Sabelhaus2017a}.

\begin{align}
\displaystyle\min_{U_{t\rightarrow t+N|t}} & \sum_{k = 0}^{N} J( \xi_{t+k|t}, u_{t+k|t}, \xi^{ref}_{t+k}) \label{eq:opt_main}\\
    & \text{subj. to:} \notag \\
    & \xi_{t+k+1} = A_{t} \xi_{t+k} + B_{t} u_{t+k} + c_{t} \label{eq:sys_dynamics}\\
    & u_{min} \leq u_{t+k} \leq u_{max} \label{eq:u_lim}\\
    & \|u_{t} - u_{(t-1)}\|_\infty \leq w_1 \label{eq:smoothing_u1}\\
    & \|u_{t+k} - u_{t}\|_\infty \leq w_2, \; k=1..(N-1) \label{eq:smoothing_uk}\\
    & \|u_{t+N} - u_{t}\|_\infty \leq w_3 \label{eq:smoothing_uN}\\
    & \|\xi(1:6)_{t+k} - \xi(1:6)_{t+k-1}\|_\infty \leq w_4 \label{eq:smoothing_x1}\\
    & \|\xi(13:18)_{t+k} - \xi(13:18)_{t+k-1}\|_\infty \leq w_5 \label{eq:smoothing_x2}\\
    & \|\xi(25:30)_{t+k} - \xi(25:30)_{t+k-1}\|_\infty \leq w_6 \label{eq:smoothing_x3}\\
    & \xi(3)_{t+k} + w_7 \leq \xi(15)_{t+k} \label{eq:collision_2}\\
    & \xi(15)_{t+k} + w_7 \leq \xi(27)_{t+k} \label{eq:collision_3}
\end{align}

Here, $N=10$ is the horizon length (a scalar), $w_1...w_7$ are constant scalar weights, and $\xi(i)_{t+k}$ denotes the $i$-th element of the state vector at time $t+k$ as predicted at time $t$.
The dynamics constraint, (\ref{eq:sys_dynamics}), consists of the linearized and discretized system at time $t$, calculated as

\[
A_t = \frac{\partial g(\xi,u)}{\partial \xi} \Bigr|_{ \substack{ \xi = \xi_{t} \\ u = u_{t-1} }}
\quad \quad
B_t = \frac{\partial g(\xi,u)}{\partial u} \Bigr|_{ \substack{ \xi = \xi_{t} \\ u = u_{t-1} }}
\]

\vspace{-0.3cm}
\[
c_t = g(\xi_t, u_{(t-1)}) - A_t \xi_t - B_t u_{(t-1)}
\]

The discretization occurs as $A_t$, $B_t$, and $c_t$ are calculated, via a simple finite-difference Euler discretization, with $k=0.001$, the same as the timestep for $t$.
At each timestep of the simulation, these matrices are calculated by numerically differentiating the equations of motion in MATLAB: the dynamics are forward simulated in each direction, and a finite-difference approximation is taken.
This approach was chosen due to computational issues with calculating additional analytical derivatives of the dynamics.

This linearization was calculated at each timestep $t$ and used for the optimization over the entire horizon, thus the notation $A_t, B_t, c_t$.
Since no trajectory of inputs was available, linearizations used the prior state's input $u_{t-1}$ instead.
For the start of the simulation, $u_{0}=\mathbf{0}$ was used.
Note that since these linearizations are not at equilibrium points, the linear system is affine, with $c_t$ being a constant vector offset.

The remaining constraints used have the following interpretations.
Constraint (\ref{eq:u_lim}) is a bound on the inputs, limiting the length of the cable rest lengths, with $u_{min},u_{max} \in \mathbb{R}^{24}$ but having the same value for all inputs.
This is the constraint that helps prevent the system from operating in the slack-cable regime, thus keeping it in one set of continous dynamics instead of behaving as a hybrid system.
Constraints (\ref{eq:smoothing_u1}), (\ref{eq:smoothing_uk}), and (\ref{eq:smoothing_uN}) are smoothing terms on the inputs, which help with the lack of an input reference trajectory.
Here $u_{(t-1)}$ is the most recent input at the start of the CFTOC problem.
Constraints (\ref{eq:smoothing_x1}), (\ref{eq:smoothing_x2}), and (\ref{eq:smoothing_x3}) are smoothing terms on the states, limiting the deviation between successive states in the trajectory.
These are needed to reduce linearization error, and are split so that the positions and angles of each vertebra could be weighted differently.
Note from (\ref{eq:smoothing_x1}-\ref{eq:smoothing_x3}) that no velocity terms are constrained.
Finally, noting that states $\xi(3)$, $\xi(15)$, and $\xi(27)$ are the z-positions of each vertebra, constraints (\ref{eq:collision_2}) and (\ref{eq:collision_3}) prevent the collision between adjacent vertebrae.

The cost function $J$, written with arbitrary time index $j$,

\vspace{-0.2cm}
\begin{align}
\begin{split}
  \displaystyle & J( \xi_{j}, u_{j}, \xi^{ref}_{j}) = \\
  & \quad \quad (\xi_{j} - \xi^{ref}_{j})^\top Q^j (\xi_{j} - \xi^{ref}_{j}) \; +\\
  & \quad \quad (\xi_{j} - \xi_{(j-1)})^\top S^j (\xi_{j} - \xi_{(j-1)}) \; + \\
  & \quad \quad w_8  \lVert (u_{j} - u_{(j-1)}) \rVert_{\infty} \label{eq:obj_fun}\\
\end{split}
\end{align}


As before, $w_8$ is a scalar, while $Q$ and $S$ are constant diagonal weighting matrices.
Here, $Q$ penalizes the tracking error in the states, $S$ penalizes the deviation in the states at one timestep to the next, and $w_8$ penalizes the deviations in the inputs from one timestep to the next.
These matrices are diagonal, with blocks corresponding to the Cartesian and Euler angle dimensions, with zeros for all velocity states, according to vertebra.
Nonzeros are at states $\xi_1...\xi_6$, $\xi_{13}...\xi_{18}$, and $\xi_{25}...\xi_{30}$, recalling that $\xi \in \mathbb{R}^{36}$.
Raising each diagonal element to the power $j$ puts a heavier penality on terms farther away on the horizon.
These are defined as:

\vspace{-0.2cm}
\begin{align}
& Q_{j} = diag( w_9, \: w_9, \: w_9 \: | \: w_{10}, \: w_{10}, \: w_{10} \: | \: 0 ... 0) \in \mathbb{R}^{12 \times 12} \notag\\
& S_j = diag( w_{11}, \: w_{11}, \: w_{11} \: | \: w_{11}, \: w_{11}, \: w_{11} \: | 0...0) \in \mathbb{R}^{12 \times 12} \notag\\
& Q = diag(Q_1, \: Q_2, \: Q_3), \quad S = diag( S_1, \: S_2, \: S_3) \notag
\end{align}

The paper \cite{Sabelhaus2017a} provides more details about the simulation for this first controller.

\subsection{Controller with a reference input, without smoothing}

The above controller required significant tuning in order to get convergence. 
So, a controller was developed that could be run without as much tuning.
One way to do so is to include a reference input trajectory, so that the optimization problem had a solution which may theoretically lead to zero error.

For this second controller, a reduced-order model of the system was used.
Specifically, only one moving vertebra was simulated, and only two-dimensional dynamics were considered.
This eliminated the number of compounding variables as the controller was developed.

The optimization problem for the newer controller is:

\begin{align}
\displaystyle\min_{U_{t \rightarrow t+N|t}} \quad & p(\xi_{t+N|t}) + \sum_{k = 0}^{N-1} q(\xi_{t+k|t}, u_{t+k|t})  \label{eq:opt_main}\\
\text{s.t.} \quad & \xi_{t+k+1|t} = A_t \xi_{t+k|t} + B_t u_{t+k|t} + c_t \label{eq:sys_dynamics}\\
    & u_{min} \leq u_{t+k|t} \leq u_{max} \label{eq:u_lim}\\
    & \xi_{t+k|t}^{(2)} \geq \frac{h}{2} \label{eq:collision}\\
    & \xi_{t|t} = \xi(t) \label{eq:initialcondition}
\end{align}

The objective function components are quadratic weights of the tracking errors on both state and input:

\vspace{-0.2cm}

\begin{align}
& p(\xi_{t+N|t}) = (\xi_{t+N|t} - \xi_{t+N}^{ref})^\top P (\xi_{t+N|t} - \xi_{t+N}^{ref}) \\
& q(\xi_{t+k|t}, u_{t+k|t}) = \notag \\
& \quad \quad (\xi_{t+k|t} - \xi_{t+k}^{ref})^\top Q (\xi_{t+k|t} - \xi_{t+k}^{ref}) \\
& \quad \quad (u_{t+k|t} - u_{t+k}^{ref})^\top R (u_{t+k|t} - u_{t+k}^{ref}) \notag
\end{align}

The reference input trajectory, $u^{ref}$, was calculated using the inverse kinematics for the positions (and assuming zero velocity) of a specifc reference state $\xi^{ref}$. 
Those inverse kinematics follow the algorithm which is used in \cite{friesen2014,Sabelhaus2015a,Schek1974}.

The constraints have the following interpretation.
Constraint (\ref{eq:u_lim}) is a box constraint on the inputs so that the springs cannot have negative length (violating the dynamics assumptions), and also cannot become too large (where the dynamics solution also becomes unrealistic.)
Constraint (\ref{eq:collision}) denotes a minimum bound on the second element in the state, the $z$-position, which prevents collision between the moving vertebra and the static vertebra, where the vertebrae each have height $h$.
Finally, constraint (\ref{eq:initialcondition}) assigns the initial condition at the starting time of the CFTOC.

The constants used in this optimization are
\begin{equation}
u_{min}=0, \quad u_{max}=0.3, \quad N=4, \quad h=0.15,
\end{equation}
and the objective function weights are $Q=P=I$ and $R=2I$.

\section{RESULTS}

\begin{figure}[thpb]
    \centering
    \includegraphics[width=1\columnwidth]{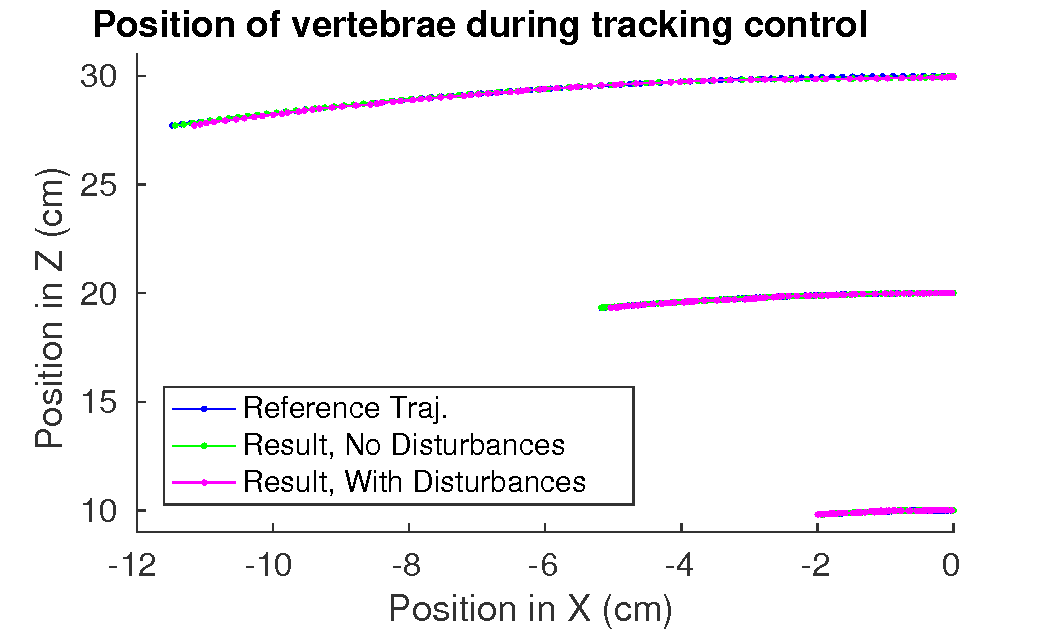}
    \caption{Positions in the $X$-$Z$ plane of all 3 of the vertebrae, including the reference and the two simulations (with/without disturbances), as the robot performs a counterclockwise bend. Blue trajectories are same as those in Fig. 1.}
    \label{fig:allvert}
\end{figure}

\begin{figure}[thpb]
    \centering
    \includegraphics[width=0.9\columnwidth]{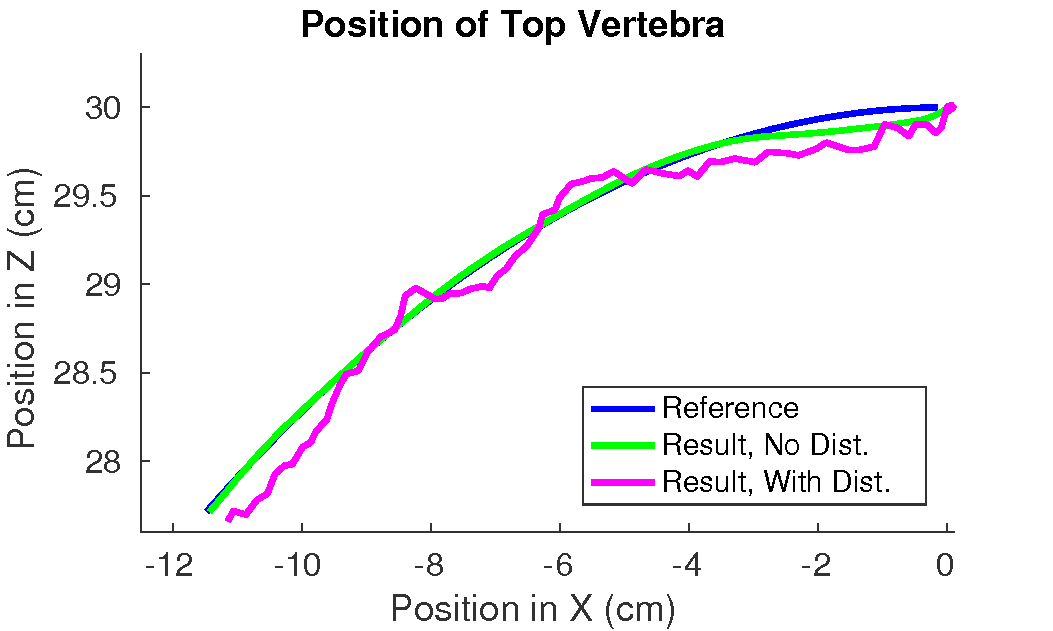}
    \caption{Positions in the X-Z plane of the top vertebra, including the reference and the two simulations (with/without disturbances), as the robot performs a counterclockwise bend. The vertebra tracks the trajectory closely.}
    \label{fig:topvert}
    \vspace{-0.3cm}
\end{figure}

\subsection{Controller without reference input}

The first controller tracked the positions of the vertebrae with extremely low error, after an initial transient response.
Fig. \ref{fig:allvert} shows the paths of the vertebrae in the $X$-$Z$ plane as they sweep through their counterclockwise bending motion, including the reference trajectory (blue), the resulting trajectory with MPC controller and no disturbances (green), and the result of the controller with disturbances (magenta).
Fig. \ref{fig:topvert} shows a zoomed-in view of the top vertebra, which had the largest tracking errors of the three vertebrae.

\subsection{Controller with reference input}

\begin{figure}[thpb]
    \centering
    \includegraphics[width=0.9\columnwidth]{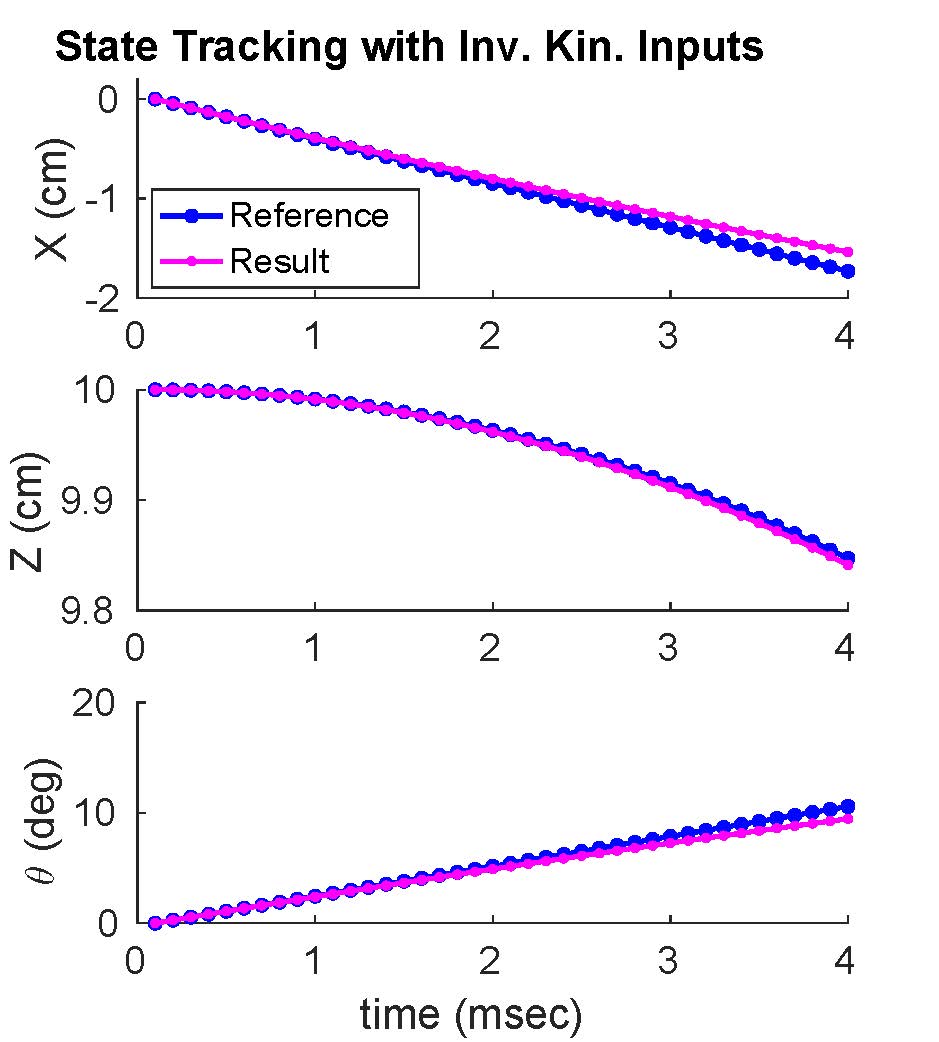}
    \caption{Positions and rotations of the single vertebra for the controller on the reduced-order model of the spine. Though this controller does not converge like the other, it does not require the extensive tuning of the other controller.}
    \label{fig:ref_states}
\end{figure}

\begin{figure}[thpb]
    \centering
    \includegraphics[width=0.9\columnwidth]{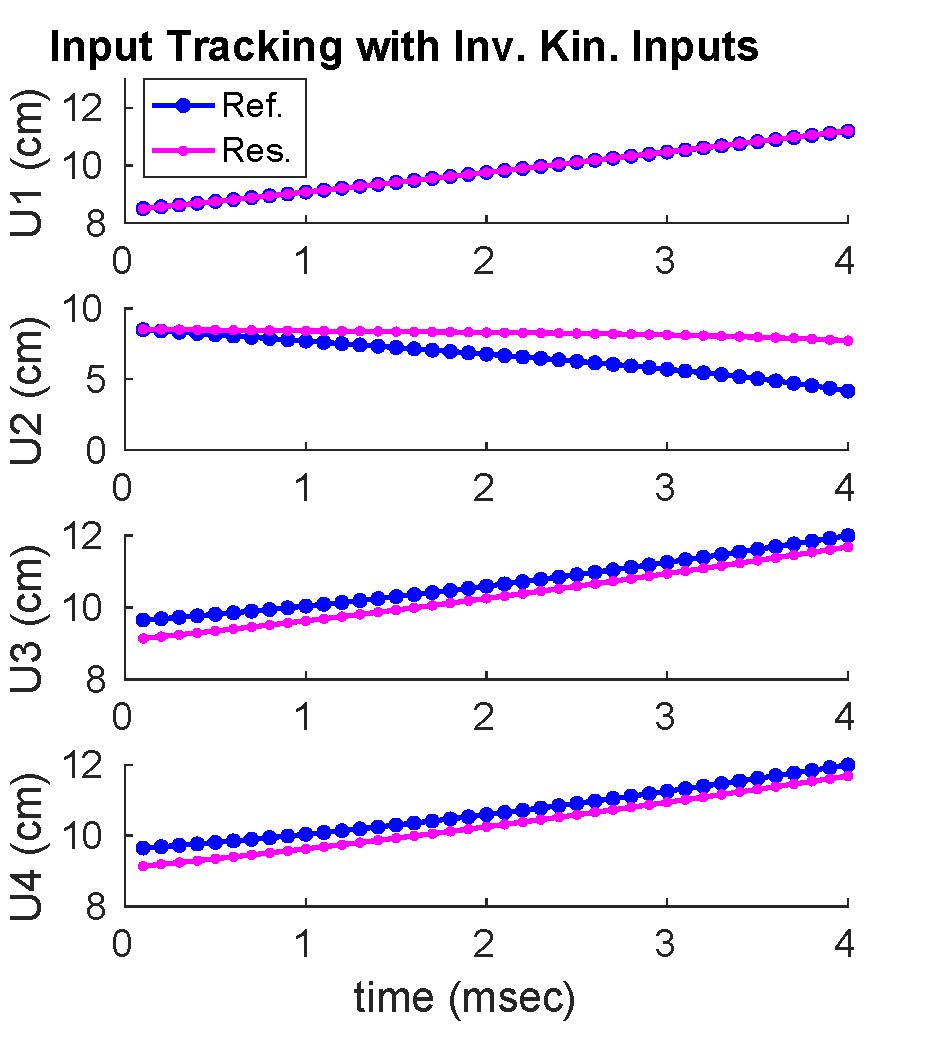}
    \caption{Input tracking for the same controller as in Fig. \ref{fig:ref_states}. Similarly, though the errors are nonzero, they are still reasonable. Considering that these inputs are for a pseudo-static spine, whereas this spine was moving relatively quickly, these results are expected.}
    \label{fig:ref_inputs}
\end{figure}

The controller for the two-dimensional, single-vertebra spine with reference input tracking does not currently perform as well as the controller with hand-tuned smoothing constraints.
However, it does not go unstable and fail, as does a controller without either smoothing or input tracking.
Figure \ref{fig:ref_states} shows the tracking of the single vertebra for each of its three kinematic states ($X$, $Z$, and angle $\theta$), showing good tracking performance.
Figure \ref{fig:ref_inputs} shows the input reference for the four cables in this system for the same simulation.
These results show promise for future work, once additional possible complications are resolved relating to discretization errors and speed of tracking.

\section*{ACKNOWLEDGEMENTS}

This research was supported by NASA Space Technology Research Fellowship no. NNX15AQ55H.





\bibliographystyle{IEEEtran}
\bibliography{IEEEabrv,bibliographies/library,bibliographies/abishek_bib}

\end{document}